\documentclass[12pt,a4paper]{article}
\usepackage{jheppub}

\allowdisplaybreaks[4]
\usepackage{amsmath,bm}
\usepackage{subfigure}

\usepackage{physics}
\usepackage{amsfonts}
\usepackage{mathtools}
\usepackage{color}
\usepackage{amsthm,amssymb}

\newcommand{\pa}{\partial}
\newcommand{\al}{\alpha}
\newcommand{\be}{\beta}
\newcommand{\ga}{\gamma}

\newcommand{\del}{\delta}

\newcommand*{\DAl}{\!\mathop{}\mathbin{\Box}}
\newcommand{\eff}{\text{eff}}

\title{
Kaluza-Klein monopole with scalar multiplet hair
}

\author[1]{Takaaki Ishii,}
\author[2]{Keiju Murata,}
\author[2]{Keita Sugawara}
\affiliation[1]{Department of Physics, College of Science, Rikkyo University, Nishi-Ikebukuro, Tokyo 171-8501, Japan}
\affiliation[2]{Department of Physics, College of Humanities and Sciences, Nihon University, Sakurajosui, Tokyo 156-8550, Japan}
\emailAdd{ishiitk@rikkyo.ac.jp}
\emailAdd{murata.keiju@nihon-u.ac.jp}
\emailAdd{chkt23002@g.nihon-u.ac.jp}

\abstract{We construct Kaluza-Klein monopole solutions with scalar hair provied by a massive complex scalar field multiplet that minimally couples to five-dimensional Einstein gravity. Writing the scalar field multiplet in terms of the Wigner D-matrices, we introduce the ansatz of the scalar multiplet compatible with the symmetries of the Gross-Perry-Sorkin monopole, on which the scalar hair grows. We give the ansatz for a multiplet with arbitrary number of components, whereas we show numerical solutions of the hairy Kaluza-Klein monopole specifically for the cases of scalar triplet and quadruplet. These generaize the preceding study on a doublet \cite{Brihaye:2023vox}. We find that the range of the mass and angular momentum of the hairy solutions are larger for higher multiplets.}

\preprint{RUP-25-18}

\begin{document}
\maketitle

\section{Introduction}

Boson stars are compact gravitational configurations formed by localized scalar fields. While their construction has been a long-standing topic in general relativity, considerable efforts have also been made to explore scalar hair around black holes. Considering higher dimensions has the advantage of simplification that even for rotating solutions one can use cohomogeneity-1 metric, i.e.~the metric components depend only on one variable. Imposing the ansatz consistent with the cohomogeneity-1 metric on the scalar field is an efficient strategy. Considering a complex scalar field doublet in five dimensions (despite the presence of a self-interacting potential), Ref.~\cite{Hartmann:2010pm} constructed cohomogeneity-1 rotating boson stars. These avoid the assumptions of the no scalar hair theorem \cite{Bekenstein:1996pn}. Cohomogeneity-1 boson stars and Myers-Perry black holes with scalar hair were then constructed for a massive scalar field without self-interaction \cite{Brihaye:2014nba}. Using the cohomogeneity-1 form makes construction much easier than cohomogeneity-2 rotating hairy black holes \cite{Herdeiro:2014goa,Herdeiro:2015kha}.

The Kaluza-Klein theory \cite{Kaluza:1921tu,Klein:1926tv} is a simple extension of four to five dimensions by adding a compact extra direction. The Kaluza-Klein mass is expected to trigger the emergence of scalar hair. While the original Kaluza-Klein spacetime is too simple, there is an interesting vacuum solution called the Gross-Perry-Sorkin (GPS) monopole \cite{Gross:1983hb,Sorkin:1983ns}. The good thing is that the GPS-monopole metric is cohomogeneity-1. In the presence of a massive complex scalar field doublet, scalar hair has been introduced in the GPS monopole as well as its black hole counterpart \cite{Brihaye:2023vox}.

Aside from spacetimes without cosmological constant, there is development in constructing the same class of solutions in asymptotically anti-de Sitter (AdS) spacetime, which has a negative cosmological constant and whose asymptotic spacetime structure helps to accommodate scalar hair. For a complex scalar field doublet, five-dimensional cohomogeneity-1 AdS boson stars and hairy black holes were obtained \cite{Dias:2011at}. Interestingly, non-stationary solutions with ``gravitational hair'', called black resonators, were also obtained in AdS spacetime first in four dimensions \cite{Dias:2015rxy}, and then cohomogeneity-1 black resonators were found in five dimensions \cite{Ishii:2018oms}, as well as related resonating configurations \cite{Ishii:2019wfs,Garbiso:2020dys}. In studies in the AdS spacetime, it has been pointed out that the scalar hair ansatz of the complex scalar doublet can be generalized to higher multiplets by using the Wigner D-matrices while keeping the cohomogeneity-1 metric \cite{Ishii:2020muv,Ishii:2021xmn}. This idea can be applied to scalars other than in the AdS spacetime. Meanwhile, in the absence of the cosmological constant, these are also six-dimensional black strings with gravitational hair \cite{Dias:2022mde,Dias:2022str,Dias:2023nbj}, where the Kaluza-Klein momentum in the direction of the string triggers gravitational hair, but scalar hair did not grow in the simple Kaluza-Klein spacetime.

In this paper, we construct Kaluza-Klein monopoles with scalar hair provided by a scalar multiplet. Specifically, we introduce the scalar hair that preserves the symmetries of the GPS-monopole metric. To this end, we express the components of the scalar field multiplet in terms of the Wigner D-matrices and introduce the ansatz of the scalar field by using the methodology developed in asymptotically AdS spacetime in \cite{Ishii:2020muv,Ishii:2021xmn}. This gives a general description of the previous construction for a complex scalar field doublet \cite{Brihaye:2023vox}. We call such a solution a hairy Kaluza-Klein monopole.\footnote{While it is a straightforward generalization also to find solutions with black hole horizons, here we focus only on horizonless monopole solutions for brevity.} While we write the ansatz for a multiplet with an arbitrary number of components, we show numerical results for the cases with a few components. We discuss the mass and angular momentum of the hairy Kaluza-Klein monopoles.

The hairy Kaluza-Kelin monopole is rotating, and the complex phase of the scalar field also is time dependent. For this class of hairy solutions, it is often argued that the components of the complex scalar field have a harmonic time dependence $\propto e^{-i \omega t}$. In this paper, by a gauge freedom to change coordinate frames, we use the rotating frame at infinity, in which the scalar field is time independent but the spacetime is asymptotically rotating. Undoing the coordinate change to go back to the non-rotating frame at infinity, we can retain the harmonic time dependence.

The structure of this paper is as follows. In Sec.~\ref{GPSiv}, we review the vacuum GPS-monopole metric and examine its isometries. In Sec.~\ref{Setup}, we introduce the setup for constructing hairy Kaluza-Klein monopoles in Einstein gravity with a complex scalar multiplet. After explaining the numerical methods in Sec.~\ref{Method}, we show the main results in Sec.~\ref{result}. Sec.~\ref{Conclusion} is the conclusion and discussion. Appendix~\ref{Counterterm} provides details of obtaining conserved quantities.

\section{Gross-Perry-Sorkin monopole}\label{GPSiv}

The Gross-Perry-Sorkin (GPS) monopole is an exact solution of the $5$-dimensional vacuum Einstein equation $G_{\mu\nu}=0$. 
The metric of the GPS monopole is given by
\begin{equation}
    ds^2=-dt^2+\qty(1+\frac{2N}{\rho})d\rho^2+\rho^2\qty(1+\frac{2N}{\rho})(\sigma^2_1+\sigma^2_2)+\frac{4N^2}{1+\frac{2N}{\rho}}\sigma^2_3\ ,
    \label{gps}
\end{equation}
where $N$ is the NUT parameter and $\sigma_i$ $(i=1,2,3)$ are the 1-forms defined by\footnote{By dimensional imensional reduction in the $\chi$ direction, the off digonal $(\chi,\theta)$ component of the metric from $\sigma_3^2=(d\chi+\cos{\theta}d\theta)^2$ turns into the gauge potential $A_\phi=\cos{\theta}$ in the effective four dimensional picture. This is the reason why the solution is called a monopole~\cite{Gross:1983hb,Sorkin:1983ns}.} 
\begin{equation}
    \begin{split}
    \sigma_1&=-\sin{\chi}d\theta+\cos{\chi}\sin{\theta}d\phi\ ,\\
    \sigma_2&=\cos{\chi}d\theta+\sin{\chi}\sin{\theta}d\phi\ ,\\
    \sigma_3&=d\chi+\cos{\theta}d\phi\ .
    \end{split}
    \label{sigmas}
\end{equation}
The angular coordinates~($\theta,\phi,\chi$) have the coordinate ranges $0\leq\theta\leq\pi,0\leq\phi<2\pi$, and $0\leq\chi<4\pi$.
The metric of the round $S^3$ with unit radius is described by these $1$-forms as $d\Omega^2_3=(\sigma^2_1+\sigma^2_2+\sigma^2_3)/4$.
We will shortly see that $\sigma_i$ are invariant under $SU(2)$-transformation. 
The GPS monopole is a one-parameter family characterized by $N$, which is called the NUT parameter since the spatial part of the GPS monopole has the form of the four-dimensional NUT instanton \cite{Hawking:1976jb}.

Near infinity $\rho\sim \infty$, the GPS-monopole metric becomes
\begin{equation}
\begin{split}
    ds^2&\simeq -dt^2+d\rho^2+\rho^2(\sigma_1^2+\sigma_2^2) +4N^2 \sigma_3^2\\
    &= -dt^2+d\rho^2+\rho^2(d\theta^2+\sin^2\theta d\phi^2) + 4N^2 (d\chi+\cos{\theta}d\phi)^2\ .
    \end{split}
    \label{vinf}
\end{equation}
This has the form of the twisted $S^1$ bundle over four-dimensional Minkowski spacetime. The topology of spacetime is asymptotically $R^{3,1} \times S^1$, and the NUT parameter $N$ corresponds to the compactification scale at infinity.

The coordinate origin $\rho=0$ is a coordinate singularity in the $\rho$ coordinate. To understand the structure of this spacetime around the origin $\rho=0$, let us introduce a new radial coordinate $r$ as 
\begin{equation}
    \frac{r^2}{4}=\rho^2(1+\frac{2N}{\rho})\ .
    \label{r_to_rho}
\end{equation}
Then, the GPS-monopole metric is rewritten as 
\begin{multline}
    ds^2=-dt^2+\frac{(\sqrt{r^2+4N^2}+2N)^2}{4(r^2+4N^2)}dr^2\\+\frac{r^2}{4}\left[\sigma^2_1+\sigma^2_2+\frac{16N^2}{(\sqrt{r^2+4N^2}+2N)^2}\sigma^2_3\right] \ .
    \label{cgGPS}
\end{multline}
Around the origin $r\sim 0$, the above metric is manifestly regular and behaves as the five-dimensional Minkowski spacetime,
\begin{equation}
    ds^2\simeq -dt^2+dr^2+\frac{r^2}{4}\qty(\sigma^2_1+\sigma^2_2+\sigma^2_3) = -dt^2+dr^2+r^2d\Omega_3^2\ .
    \label{v0}
\end{equation}

\begin{figure}[t]
    \centering
    \includegraphics[scale=0.7]{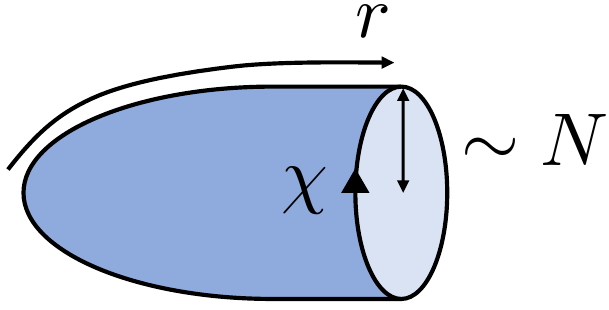}
    \caption{
   A schematic diagram of the GPS-monopole metric. In $(r,\chi)$-coordinates, the spacetime is cylindrical at infinity and locally flat around the cooridnate origin. The coordinate $\chi$ is compactified with radius $\sim N$ at infinity.
    }
    \label{metric}
\end{figure}

Thus, the GPS-monopole spacetime is schematically considered to be cigar-shaped, as shown in Fig.~\ref{metric}. 

To consider the symmetries of the GKP monopole spacetime, let us introduce $SU(2)$ generators that satisfy $[\xi_i, \xi_j] = \epsilon_{ijk} \xi_k$ $(i, j, k = 1,2,3)$ as follows:
\begin{equation}
    \begin{split}
        \xi_1&=\cos{\phi}\partial_\theta+\frac{\sin{\phi}}{\sin{\theta}}\partial_\chi-\cot{\theta}\sin{\phi}\partial_\phi\ , \\
    \xi_2&=-\sin{\phi}\partial_\theta+\frac{\cos{\phi}}{\sin{\theta}}\partial_\chi-\cot{\theta}\cos{\phi}\partial_\phi\ ,\\
    \xi_3&=\partial_\phi\ . 
    \end{split}
\end{equation}
One can check that the 1-forms defined in Eq.~\eqref{sigmas} are invariant under $SU(2)$-transformations: $\mathcal{L}_{\xi_j} \sigma_i = 0$, where $\mathcal{L}$ denotes the Lie derivative. It follows that these generators are Killing vectors of the GPS-monopole spacetime. The $1$-forms $\sigma_i$~($i=1,2,3$) satisfy the Mauer-Cartan equations for $SU(2)$: $d\sigma_i=(1/2)\epsilon_{ijk}\sigma_j\wedge\sigma_k$. In addition to the above $SU(2)$ symmetry, the GKP monopole spacetime is also invariant under the $U(1)$ ``rotation'' of $(\sigma_1, \sigma_2)$: 
$\sigma_1 + i \sigma_2 \to e^{i\lambda} (\sigma_1 + i \sigma_2)$ ($\lambda \in \mathbb{R}$), under which the combination $\sigma_1^2+\sigma_2^2$ in the metric is invariant.
This transformation is generated by $\partial_\chi$. In addition, the GPS-monopole spacetime is independent of time, and the time translation $\partial_t$ is also a Killing vector.  In summary, the isometry group of the GPS-monopole spacetime is given by $R_t \times SU(2) \times U(1) \simeq R_t \times U(2)$.

\section{Setup for constructing hairy Kalza-Klein monopoles}\label{Setup}
\subsection{Metric ansatz}

In this paper, we introduce scalar hair in the GPS-monopole spacetime in a way that preserves the aforementioned $R_t \times U(2)$ isometries.
We start with the action of the Einstein-complex scalar field system in $5$-dimensions given by  
\begin{equation}  
    S = \frac{1}{16\pi G_5} \int d^5x \sqrt{-g} \left( R - g^{\alpha\beta} \partial_\alpha \bm{\Pi}^* \cdot \partial_\beta \bm{\Pi} - \mu^2 \bm{\Pi}^* \cdot \bm{\Pi} \right)\ ,  
    \label{L5}  
\end{equation}  
where $\bm{\Pi}$ is an $n$-component complex scalar multiplet, and $\mu$ is the mass of the scalar field. The integer $n$ will be specified later.

The equations of motion of this system are given by
\begin{equation}
    G_{\mu\nu}=T_{\mu\nu}\ ,\quad \DAl\bm{\Pi}=\mu^2\bm{\Pi}\ ,
    \label{eom}
\end{equation}
where the energy momentum tensor takes the form
\begin{equation}
    T_{\mu\nu}=\pa_{(\mu}\bm{\Pi}^*\cdot\pa_{\nu)}\bm{\Pi}-\frac{1}{2}g_{\mu\nu}\qty{g^{\al\be}\pa_{(\al} \bm{\Pi}^*\cdot\pa_{\be)} \bm{\Pi}+\mu^2\bm{\Pi}^*\cdot\bm{\Pi}}\ .
\end{equation}
We assume that the spacetime has the same isometry group as the GPS monopole. 
The metric ansatz is
\begin{equation}
    ds^2=-f(r) dt^2+\frac{dr^2}{g(r)}+\frac{r^2}{4}[\sigma_1^2+\sigma_2^2+\beta(r)(\sigma_3+2h(r)dt)^2]\ .
    \label{ansatz}
\end{equation}
The GPS-monopole solution \eqref{cgGPS} corresponds to choosing
\begin{equation}
\begin{split}
    &f(r)=1\ ,\quad 
    g(r)=\frac{4(r^2+4N^2)}{(\sqrt{r^2+4N^2}+2N)^2}\ ,\\
    &\beta(r)=\frac{16N^2}{(\sqrt{r^2+4N^2}+2N)^2}\ ,\quad h(r)=0\ ,\quad \bm{\Pi}=0\ .
\end{split}
\label{GPSsol}
\end{equation}

Even if the spacetime has the $R_t \times U(2)$ symmetries as mentioned above, the scalar field does not necessarily have to share the same symmetry. This is because the condition imposed by the spacetime symmetry is $\mathcal{L}_{\xi_A}T_{\mu\nu}=0$, but not $\mathcal{L}_{\xi_A}\bm{\Pi}=0$, where $\xi_A$ is a Killing vector, i.e. $\xi_A=(\partial_t, \partial_\chi,\xi_i)$. The energy-momentum tensor is quadratic in the scalar field, so if the scalar field transforms covariantly under a symmetry transformation, then the energy-momentum tensor can remain invariant. In the following, we will write the scalar field using the mode functions that transform covariantly under $SU(2)$. Employing the scalar field proportional to these functions, we can find the configuration of the scalar hair that preserves the symmetries of the spacetime.

\subsection{Wigner D-matrices}\label{WD}

The Wigner D-matrices are defined as eigenfunctions of generators of $SU(2)\times U(1)$ as
\begin{equation}
    L^2 D^j_{~m,k}=j(j+1)D^j_{~m,k}\ ,\quad 
    L_3 D^j_{~m,k}=m D^j_{~m,k}\ ,\quad 
    R_3 D^j_{~m,k}=k D^j_{~m,k}\ ,
    \label{evf}
\end{equation}
where we defined the angumar momentum operators as 
\begin{equation}
    L_i = i\xi_i\ ,\qquad R_3=i\pa_\chi\ .
    \label{amo}
\end{equation}
The Casimir operator of $SU(2)$ is given by $L^2=L_1^2+L_2^2+L_3^3$. (We follow the convention of the Wigner D-matrices in \cite{Hu:1974hh}, see also \cite{Ishii:2021xmn}.) 
The Wigner D-matrices are labeled by three ``quantum numbers'' $(j,m,k)$. These take values in the following ranges.
\begin{equation}
        j=0,~\frac{1}{2},~1,~\frac{3}{2},\cdots\ ,\quad 
        m=-j,~-j+1,~\cdots,j\ ,\quad 
        k=-j,~-j+1,~\cdots,j\ .
        \label{jmk_quantum_numbers}
\end{equation}

We define the following $(2j+1)$-component multiplet of $SU(2)$,
\begin{equation}
    \bm D_k=
    \begin{pmatrix}
        D^j_{~m=j,k}\\
        D^j_{~m=j-1,k}\\
        \vdots\\
        D^j_{~m=-j,k}
    \end{pmatrix} \ .
\end{equation}
Although the multiplet $\bm{D}_k$ also depends on the quantum number $j$, we will suppress it for notational simplicity. This will not cause confusion because $j$ is kept fixed. 
The multiplet $\bm{D}_k$ satisfies the orthonormality relation: 
\begin{equation}
    \bm{D}^*_{k'} \cdot \bm{D}_k=\del_{k'k}\ .
    \label{DDorth}
\end{equation}
The Wigner D-matrices satisfy the following differential formulae:
\begin{subequations}
    \begin{align}
        \pa_\theta \bm{D}_k&=-\frac{i}{2}\qty(\epsilon_k e^{-i\chi}\bm{D}_{k-1}+\epsilon_{k+1}e^{i\chi}\bm{D}_{k+1})\ ,\\
        \pa_\phi \bm{D}_k&=-ik\cos{\theta}\bm{D}_k+\frac{1}{2}\epsilon_k\sin{\theta}e^{-i\chi}\bm{D}_{k-1}-\frac{1}{2}\epsilon_{k+1}\sin{\theta}e^{i\chi}\bm{D}_{k+1}\ ,\\
        \pa_\chi \bm{D}_k&=-ik\bm{D}_k\ ,
    \end{align}
    \label{delD}
\end{subequations}
where $\epsilon_k\equiv\sqrt{(j+k)(j-k+1)}$. See \cite{Ishii:2021xmn} for the derivation of Eqs.(\ref{DDorth}) and (\ref{delD}). 

\subsection{Ansatz for complex scalar field}
\label{ansatzscalar}

We take the ansatz that the $(2j+1)$-component complex scalar field $\bm{\Pi}$ is proportional to the Wigner D-matrix as
\begin{equation}
     \bm{\Pi}(t,r,\theta,\phi,\chi)=\Phi(r)\bm D_k(\theta,\phi,\chi)\ .
     \label{multi}
 \end{equation}
In this paper, we present numerical results only for the case of $k=j$, which is expected to be the most prone to form scalar hair among different $|k|\le j$ at a fixed $j$, but the calculations in the following can be performed without specifying $k$, and we do so for the time being.

The above scalar field does not have any explicit time dependence, but this ansatz actually contains the oscillation of the scalar field. In general, the metric component $h(r)$ asymptotically approaches a constant value $h_\infty$ at infinity, which means that the metric describes a rotating frame at infinity. Transformation into the non-rotating frame \eqref{vinf} can be achieved by a coordinate transformation $\chi_\text{non-rot} = \chi_\text{rot} + 2h_\infty t$. Under this transformation, the scalar field acquires an explicit time dependence of the form $\bm{\Pi} \propto e^{-ik\chi_\text{rot}} = e^{-ik\chi_\text{non-rot} + 2ih_\infty k t}$. This implies that the scalar field oscillates with a frequency $\omega = 2h_\infty k$ in the non-rotating frame.

This time dependence is consistent with our metric ansatz \eqref{ansatz} even in the non-rotating frame.
We can also assume that $\Phi \in \mathbb{R}$ because its complex phase can be absorbed by a constant shift of $\chi$. Then, for the scalar field in Eq.(\ref{multi}), we have 
$\bm{\Pi}^\ast \cdot \bm{\Pi} = \Phi^2$. Using the differential formulae~(\ref{delD}), we also obtain
\begin{equation}
    \partial_\mu\bm{\Pi}^\ast \cdot \partial_\nu \bm{\Pi}\,dx^\mu dx^\nu = \Phi'{}^2 dr^2 + \frac{1}{4}(\epsilon_k^2+\epsilon_{k+1}^2)\Phi^2 (\sigma_1^2+\sigma_2^2)+k^2\Phi^2\sigma_3^2\ .
\end{equation}
Thus, the energy momentum tensor is invariant under $R_t \times U(2)$ and the scalar field (\ref{multi}) is consistent with the ansatz for the metric~(\ref{ansatz}). Note that when $j=1/2$ this construction reduces to the doublet scalar considered in \cite{Herdeiro:2014goa}.

\subsection{Equations of motion}

Under the metric ansatz \eqref{ansatz}, the equations of motion \eqref{eom} reduce to the simultaneous ordinary differential equations given by
\begin{equation}
    \begin{split}
    &f'= -r^4 \frac{r g(4 f\beta' +  r^3 \beta^2 h'{}^2)    + 4f\beta ( \beta + 3 g- 4)}{g (r^6\beta)'} + \frac{4 r^6 f \beta T_{rr}}{(r^6\beta)'} \ ,\\
    &g'=\frac{r^3 g \beta h'{}^2}{6f} + \frac{r g\beta'}{6\beta}\left(\frac{f'}{f} - \frac{2}{r}\right) -\frac{2(3 g+5\beta-8)}{3r}\\
    &\hspace{3cm}+2 \frac{- r^2 \beta T_{tt} + 4 r^2 \beta h T_{t3} - 4 f\beta T_{11} +4(f-r^2 \beta h ^2)T_{33}}{3rf\beta}\ ,\\
    &h''=\frac{h'}{2} \left(\frac{f'}{f}- \frac{g'}{g} -\frac{3\beta'}{\beta} - \frac{10}{r}\right)
    +\frac{4 (2 h T_{33} - T_{t3})}{r^2 g \beta }\ ,\\
    &\beta''=-\frac{\beta'}{2}\left(\frac{f'}{f} + \frac{g'}{g} - \frac{\beta'}{\beta} + \frac{6}{r}\right) -\frac{r^2 \beta^2 h'{}^2}{f} + \frac{8\beta(\beta-1)}{r^2g}+\frac{8(\beta T_{11} - T_{33})}{r^2 g} \ ,\\
    &\Phi''=-\frac{(r^6fg\beta)'}{2(r^6fg\beta)}\Phi'
    +\frac{1}{g}\left(
    \frac{2(\epsilon_k^2+\epsilon_{k+1}^2)}{r^2} 
    +4k^2\left(\frac{1}{\beta r^2}-\frac{h^2}{f}\right)
    + \mu^2
    \right)\Phi\ ,
    \end{split}
    \label{Ef}
\end{equation}
where ${}'\equiv d/dr$. We expand the energy momentum tensor by ($dt,dr,\sigma_1,\sigma_2,\sigma_3)$-basis as $T_{\mu\nu}dx^\mu dx^\nu = T_{tt}dt^2 +T_{rr}dr^2 + T_{11} \sigma_1^2 + T_{22} \sigma_2^2 +T_{33} \sigma_3^2 + 2 T_{t3}dt \sigma_3$. The nonzero components of the energy momentum tensor are explicitly given as follows:
\begin{equation}
    \begin{split}
    &T_{tt} = (f-r^2\beta h^2)\left[\frac{g \Phi'{}^2}{2} +\left( \frac{\mu^2}{2} + \frac{\epsilon_k^2+\epsilon_{k+1}^2}{r^2} + \frac{ 2 k^2}{r^2 \beta}-\frac{2 k^2 h ^2}{f}\right)\Phi^2\right]\ ,\\
    &T_{rr}=\frac{\Phi'{}^2}{2}-\frac{1}{g}\left(
    \frac{\mu^2}{2} +\frac{\epsilon_k^2+\epsilon_{k+1}^2}{r^2} +  \frac{2 k^2}{r^2 \beta } -\frac{2 k^2 h^2}{f}   
    \right)\Phi^2\ ,\\
    &T_{11}=T_{22}=-\frac{r^2 g }{8}\Phi'{}^2 - \frac{1}{2}\left(\frac{ \mu^2 r^2 }{4} -\frac{r^2 k^2 h^2}{f}   + \frac{k^2}{\beta}\right)\Phi^2\ ,\\
    &T_{33}=-\frac{r^2 g\beta }{8}\Phi'{}^2-\frac{1}{2}\left(\frac{\mu^2 r^2 \beta }{4} + \frac{(\epsilon_k^2+\epsilon_{k+1}^2)\beta}{2} - \frac{k^2 r^2 \beta  h}{f}- k^2\right)\Phi^2\ ,\\
    &T_{t3}=-\frac{r^2 g \beta h}{4}\Phi'{}^2-h\left( \frac{\mu^2 r^2 \beta }{4}  + \frac{(\epsilon_k^2+\epsilon_{k+1}^2)\beta}{2} - \frac{k^2 r^2  \beta h^2}{f} + k^2\right)\Phi^2\ .
    \end{split}
\end{equation}
We will solve the ordinary differential equations \eqref{Ef} numerically. 

\subsection{No nomal modes in the probe scalar field limit}
\label{nonomalmode}

Let us consider the limit where the amplitude of the scalar field is infinitesimal and its backreaction is negligible. In \cite{Brihaye:2023vox}, it was shown that there are no normal modes in the GPS-monopole spacetime for the scalar doublet, which corresponds to $j = 1/2$ here. In this subsection, we show that the analysis is extended to general $j$.

We consider a probe scalar multiplet in the background GPS-monopole spacetime given by Eq.~\eqref{GPSsol}. To account for the frequency of the scalar field, we shift the angular coordinate as $\chi \to \chi + (\omega/k) t$ and set $h(r) = \omega/(2k)$. In the $\rho$-coordinate defined in Eq.~(\ref{r_to_rho}), the equation governing the scalar field is
\begin{equation}
    \frac{d^2 \Phi}{d \rho^2} + \frac{2}{\rho} \frac{d \Phi}{d \rho} - \left( \left( 1 + \frac{2 N}{\rho} \right) \nu^2 + \frac{j(j+1)}{\rho^2} + \frac{k^2}{2N \rho} \right) \Phi = 0\ ,
\end{equation}
where $\nu^2 \equiv \mu^2 + \frac{k^2}{4N^2} - \omega^2$. The analytical solution to this equation is
\begin{multline}
    \Phi = e^{-\nu \rho} \bigg[ c_1 \rho^j M \left( j + 1 + \frac{k^2}{4 \nu N} + \nu N, 2j + 2, 2 \nu \rho \right) \\
    + c_2 \rho^{-j-1} M \left( - j + \frac{k^2}{4 \nu N} + \nu N, - 2j, 2 \nu \rho \right) \bigg]\ ,
\end{multline}
where $M$ denotes Kummer's confluent hypergeometric function. Near the origin, this solution behaves as
\begin{equation}
    \Phi \simeq c_1 \rho^j + c_2 \rho^{-j-1}\ .
\end{equation}
Thus, regularity at the origin requires $c_2 = 0$. Then, at infinity, the scalar field behaves as
\begin{equation}
    \Phi \simeq c_1 \frac{\Gamma(2j + 2)}{\Gamma \left( j + 1 + \frac{k^2}{4 \nu N} + \nu N \right)} \rho^j (2\nu \rho)^{-j-1 + \frac{k^2}{4 \nu N} + \nu N} e^{\nu \rho}\ .
\end{equation}
This expression diverges at infinity, unless $c_1 = 0$. Therefore, we conclude that there are no normal modes in the GPS-monopole spacetime in general $j$.
This implies that the self-gravitating effect of the scalar field plays a crucial role for scalar hair to be added to the GPS-monopole spacetime.

\subsection{Asymptotic expansion at origin}

Near the origin $r\sim 0$, the regularity of the spacetime requires the metric components to behave as
\begin{equation}
    f(r) \to f_0\ ,\quad 
    g(r) \to 1\ ,\quad
    \beta(r)\to 1\ ,\quad
    h(r)\to h_0\ ,
\end{equation}
where $f_0$ and $h_0$ are constants. With these conditions, near the origin, the Klein-Gordon equation becomes
\begin{equation}
    \Phi''+\frac{3}{r}\Phi'-\frac{4j(j+1)}{r^2}\Phi \simeq 0\ .
\end{equation}
The behavior of the regular solution is $\Phi\sim r^{2j}$ in $r\sim 0$.
The variables can be expanded near the origin by a power series as
\begin{equation}
    \begin{split}
    &f(r)=f_0+f_2r^2+\cdot\ ,\quad g(r)=1+g_2r^2+\cdots\ ,\quad \beta(r)=1+\beta_2r^2+\cdots\ ,\\
    &h(r)=h_0+h_2r^2+\cdots\ ,\quad \Phi(r)=r^{2j}(\Phi_0+\Phi_2r^2+\cdots)\ .
    \end{split}
    \label{bc0}
\end{equation}
Substituting the above power series expansion into Eq.~\eqref{Ef}, we find the coefficients order by order as\footnote{
The reason for this case division is that the behavior of the scalar field near the origin depends on $j$. For $j=1/2$, the scalar field linearly approaches zero at the origin, which affects the second-order coefficients of the metric components.
}
\begin{equation}
\begin{split}
    & f_2=0\ , \quad 
    g_2=
    \begin{cases}
        -\beta_2-\frac{1}{3}\Phi_0^2 & (j=1/2)\\
        -\beta_2  & (j\geq 1)
    \end{cases}
    \ , \quad 
    h_2=
    \begin{cases}
        \frac{1}{6}h_0\Phi_0^2 & (j=1/2)\\
        0  & (j\geq 1)
    \end{cases}
    \ , \\
    & \Phi_2=\frac{\Phi_0}{2(j+1)} \left[-j(j + \frac{3}{2})g_2 - (k^2  + \frac{j}{2})\beta_2   - \frac{k^2 h_0^2}{f_0} + \frac{\mu^2}{4}\right]\ .
\end{split}
\label{BCin}
\end{equation}
The subleading terms in the power series expansion are determined by the four parameters $(f_0,h_0,\beta_2,\Phi_0)$ that are not fixed in the series expansion around $r=0$.
The higher-order coefficients, which are not shown here, are also determined by these parameters.

We could alternatively impose the inner boundary condition as the black hole horizon at a finite coordinate value of $r$. In this paper, we focus only on the simpler horizonless setup to discuss the dependence on the number of multiplet components of the hairy solution.

\subsection{Asymptotic expansion at infinity}

At infinity $r\sim\infty$, we assume that our spacetime has the same asymptotic structure as the GPS monopole:
\begin{equation}
    f(r)\to 1\ ,\quad  g(r)\to 4\ ,\quad h(r)\to \frac{\omega}{2k}\ ,\quad  r^2 \beta(r)\to 16N^2\ ,
\end{equation}
where $\omega$ and $N$ are positive constant. Here, we assume the rotating frame at infinity so that the whole GPS monopole spacetime is rigidly rotating. 
The NUT parameter $N$ represents the scale of compactification at infinity. 
As mentioned in section~\ref{ansatzscalar}, the parameter $\omega$ corresponds to the frequency of the scalar field.

Near infinity, the Klein-Gordon equation asymptotically takes the form  
\begin{equation}
    \Phi''+\frac{2}{r}\Phi'-\frac{\mu^2_{\text{eff}}-\omega^2}{4}\Phi \simeq 0\ ,
\end{equation}
where we define the effective four-dimensional mass that includes the contribution from the Kaluza-Klein mass as  
\begin{equation}
    \mu_{\text{eff}}=\sqrt{\mu^2+\frac{k^2}{4N^2}} \ .
    \label{mueff}
\end{equation}
The regular solution of the scalar field can be found in $\omega > \mu_{\eff}$ and behaves as $\Phi\sim \exp[-\sqrt{\mu^2_{\eff}-\omega^2}\,r/2]/r$ near infinity.
Solving the equations of motion order by order, we find the power series expansion of the variables near infinity as
\begin{equation}
    \begin{split}
    &f(r)=1+\frac{C_f}{r}+\cdots\ ,\quad g(r)=4+\frac{4(C_f + C_\beta/(16N^2))}{r}+\cdots,\\
    &\beta(r)=\frac{16N^2}{r^2}+\frac{C_\beta}{r^3}+\cdots\ ,\quad
    h(r)=\frac{\omega}{2k}+\frac{C_h}{r}+\cdots,\\
    &\Phi(r)=\frac{\Phi_{\infty}}{r}\exp\left[-\frac{r}{2}\sqrt{\mu^2_{\eff}-\omega^2}\right]+\cdots.
    \end{split}
    \label{power_series_infinity}
\end{equation}
These have six parameters $(N,\omega,C_f,C_h,C_{\beta},\Phi_{\infty})$ that are not determined by the series expansion. 
Higher-order coefficients are determined by these parameters.
From these coefficients, the mass $M_\text{tot}$ and angular momentum $J$ of this spacetime are determined as
\begin{equation}
    M_\text{tot}=-\frac{\pi}{2G_5}\qty(\frac{C_\beta}{8N}+4NC_f)\ ,\quad 
    J=-\frac{8\pi N^3 C_h}{G_5}\ .
    \label{cc}
\end{equation}
To obtain these expressions, we used the counterterm method. See Appendix~\ref{Counterterm} for details.

For the GPS monopole \eqref{GPSsol}, we have $C_f=C_h=0$ and $C_\beta=-64N^3$, and the mass and angular momentum of the GPS monopole are given by $M_\text{GPS}=4\pi N^2/G_5$ and $J_\text{GPS}=0$. We use $M$ defined as the mass measured from that of the GPS monopole,
\begin{equation}
    M=M_\text{tot}-M_\text{GPS}=-\frac{\pi}{2G_5}\qty(\frac{C_\beta}{8N}+4NC_f+8N^2)\ .
    \label{relativemass}
\end{equation}

For hairy Kaluza-Klein monopoles, we will see that the angular momentum $J$ is non-zero when there is non-trivial scalar hair.

\section{Numerical methods}\label{Method}

In the asymptotic expansions at origin and infinity, there are ten free parameters $(f_0,h_0,\beta_2,\Phi_0,N,\omega,C_f,C_h,C_{\beta},\Phi_{\infty})$. We take the units where $\mu=1$, that is, we measure dimensionful quantities in units of $\mu$. 
To obtain numerical solutions of the equations of motion, we employ the two-sided shooting method. 
We integrate the equations from the inner boundary ($r=r_\text{min}$) to the matching point ($r=r_\text{match}$).
Independently, we also integrate the equations from the outer boundary ($r=r_\text{max}$) to the matching point ($r=r_\text{match}$).
We tune the parameters so that the solutions from both regions connect smoothly at the matching point.
We take $r_\text{min}=10^{-3}$, $r_\text{match}=10$ and $r_\text{max}=10^2$ in our numerical computations.

The equations of motion~(\ref{Ef}) consist of two first-order ordinary differential equations and three second-order ordinary differential equations. Thus, a total of $2 + 3 \times 2 = 8$ matching conditions are necessary. Meanwhile, the asymptotic expansion contains 10 free parameters. Hence, two parameters must be treated as input parameters. 
We take $(N, \Phi_0)$ as input parameters. 
This means that the hairy Kalza-Klein monopole solution can be constructed as a two-parameter family of $N$ and $\Phi_0$.

In fact, the number of shooting parameters can be reduced from eight to seven by coordinate degrees of freedom.
By rescaling the time coordinate $t\to ct$, the metric components are transformed as $f\to c^2 f,~h\to ch$. 
Using this scaling, we can always match the function $f(r)$ at $r=r_\text{match}$. Using this freedom, we can set $f_0=1$. The other seven matching conditions are given by
\begin{equation}
\begin{split}
    &g_\textrm{in}=g_\text{out}\ ,\quad
    \sqrt{\frac{f_\text{out}}{f_\text{in}}}h_\textrm{in}=h_\text{out}\ ,\quad 
    \sqrt{\frac{f_\text{out}}{f_\text{in}}}h'_\textrm{in}=h'_\text{out}\ ,\\
    &\beta_\textrm{in}=\beta_\text{out}\ ,\quad 
    \beta'_\textrm{in}=\beta'_\text{out}\ ,\quad 
    \Phi_\textrm{in}=\Phi_\text{out}\ ,\quad 
    \Phi'_\textrm{in}=\Phi'_\text{out}\ ,
\end{split}
\end{equation}
where the subscripts ``in'' and ``out'' denote the inner values at $r = r_\text{match} - 0$ and the outer values at $r = r_\text{match} + 0$, respectively. Once the shooting method has converged, the numerical solution for $r < r_\text{match}$ is transformed as $f \to (f_\text{out}/f_\text{in}) f$ and $h \to \sqrt{f_\text{out}/f_\text{in}} h$, and we obtain a smooth solution over the entire space.
We define the errors of the shooting method as $\text{[error]} = |g_\text{in}-g_\text{out}|+\cdots+|\Phi'_\textrm{in}-\Phi'_\text{out}|$. We solved the solutions with an accuracy $\text{[error]}\lesssim 10^{-6}$. 

Whether the shooting method successfully converges or not strongly depends on the quality of the initial guess. For readers who wish to check the numerical calculations, we provide one set of parameters for which the shooting method has successfully converged: for $j=k=1/2$, $N=1$, $\Phi_0=0.02$, we find a solution when $h_0=1.09661426$, $\beta_2=-0.104868598$, $\omega=0.9777642290$, $C_f=-2.77504845$, $C_\beta=-40.5764592$, $C_h=-0.250462748$, $\Phi_\infty=127.473857$.
By gradually varying $N$ and $\Phi_0$, the shooting method can be continued to expand the solution space.
By formally treating the quantum number $j$ as a real number, the solution can be extended to the region $j > 1/2$.
Then, for $1/2<j<1$, we set
$g_2=-\beta_2-\Phi_0^2/3\times 2(1-j)$, $h_2=h_0\Phi_0^2/6\times 2(1-j)$ in Eq.(\ref{BCin}) so that the boundary conditions are smoothly connected at $j=1$, and we can find a good initial guess for $j=1$ from $j=1/2$.

\section{Results}\label{result}

\begin{figure}[t]
  \centering
\subfigure
 {\includegraphics[scale=0.42]{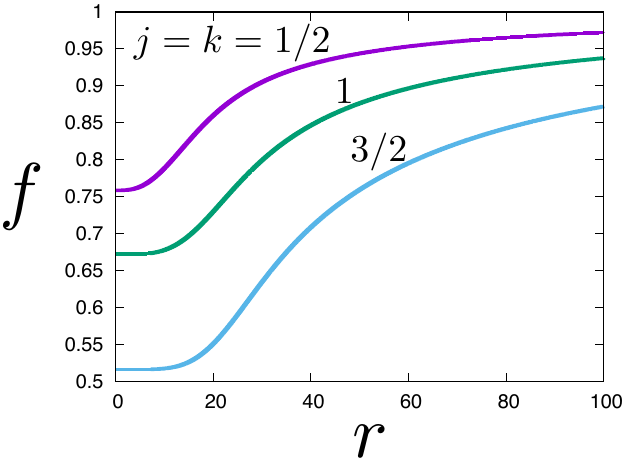}\label{f}
  }
  \subfigure
 {\includegraphics[scale=0.42]{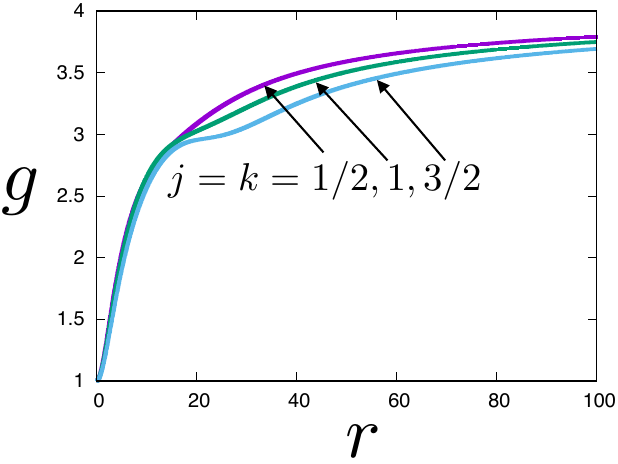}\label{g}
  }
   \subfigure
 {\includegraphics[scale=0.42]{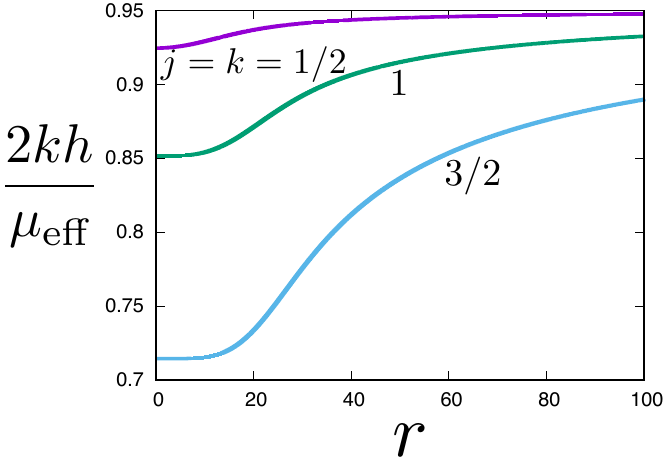}\label{h}
  }
  \subfigure
 {\includegraphics[scale=0.42]{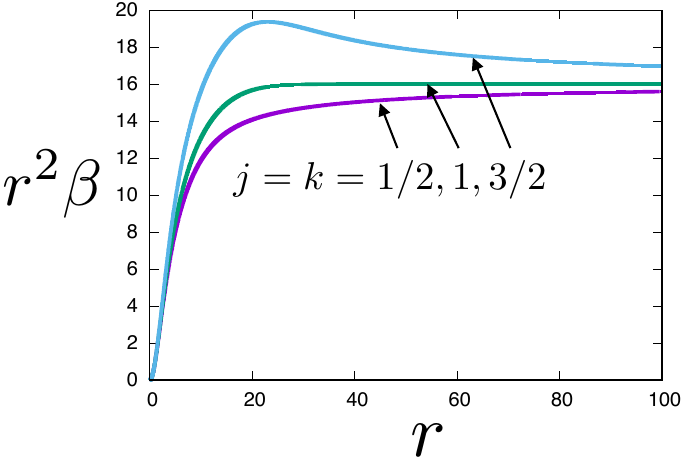}\label{beta}
  }
  \subfigure
 {\includegraphics[scale=0.42]{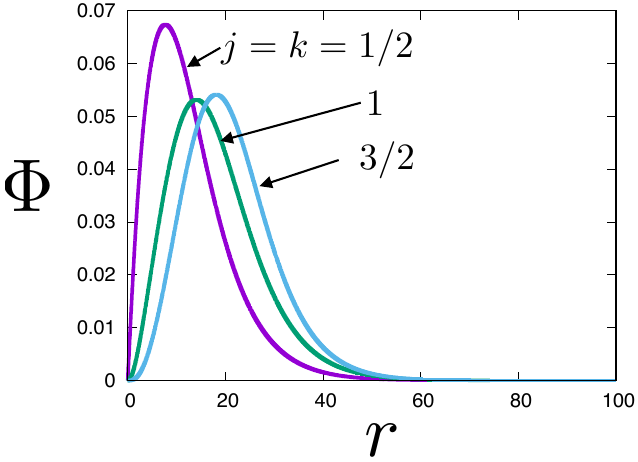}\label{phi}
  }
 \caption{Functional profiles of metric compoments $f,g,h,\beta$ and scalar field $\Phi$ for $j=k=1/2,1,3/2$. The NUT parameter and frequency of the scalar field are fixed as $N=1$ and $\omega/\mu_\text{eff}=0.95$.
}
 \label{metric_compts}
\end{figure}

Using the numerical method described in the previous section, we solved Eq.~\eqref{Ef} in the range $r_\text{min} = 10^{-3}$ and $r_\text{max} = 10^2$, with the matching point set at $r_\text{match} = 10$. In Fig.~\ref{metric_compts}, we show profiles of numerical solutions of $f(r),\,g(r),\,h(r),\,\beta(r),\,\Phi(r)$ for $j = k = 1/2, 1, 3/2$. In the figure, the input parameters are set to $N = 1$ and $\omega / \mu_{\text{eff}} = 0.95$, instead of specifying $\Phi_0$. For visibility, also, instead of plotting the functions $h$ and $\beta$ themselves, we show $2kh/\mu_\text{eff}$ and $r^2\beta$, which asymptote to $\omega/\mu_\text{eff}$ and $16N^2$ at infinity, respectively.
The solutions are smoothly connected at $r = r_\text{match}$ and are regular throughout the domain of $r$. 
The scalar field is localized around the origin but vanishes in $r \to 0$.  
This behavior arises because the scalar field behaves as $\Phi \sim r^{2j}$ near the origin and $\Phi \sim e^{-\sqrt{\mu_\text{eff}^2 - \omega^2} \, r/2}$ at infinity. Also, the location of the peak moves outward as $j$ increases. As shown in Section~\ref{nonomalmode}, in the probe limit, the GPS-monopole spacetime does not support scalar hair. However, when self-gravity is taken into account, the scalar hair contributes nonlinearly to the metric functions and becomes localized.

\begin{figure}[t]
  \centering
\subfigure[$j=k=1/2$]
 {\includegraphics[scale=0.45]{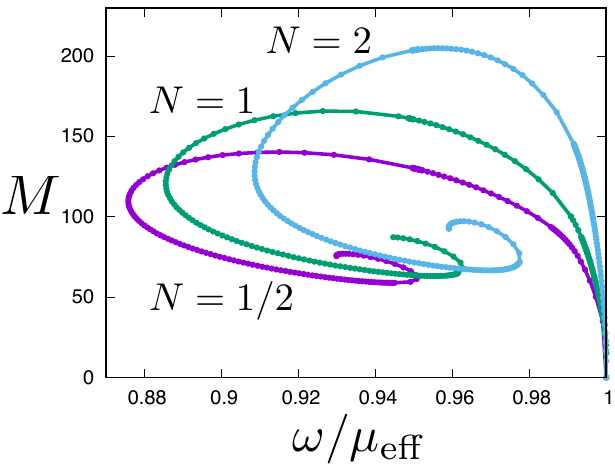}\label{Mj05}
  }
  \subfigure[$j=k=1$]
 {\includegraphics[scale=0.45]{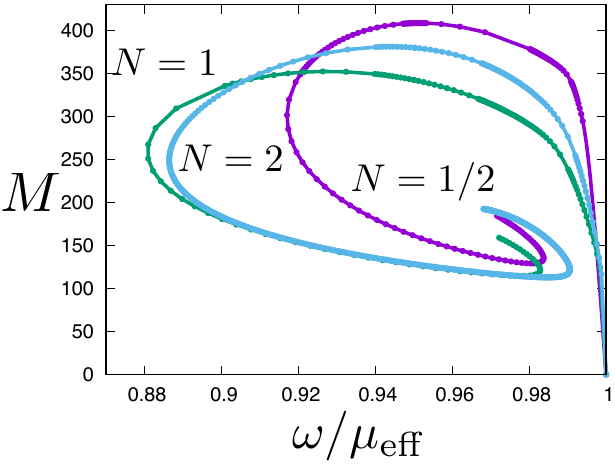}\label{Mj1}
  }
   \subfigure[$j=k=3/2$]
 {\includegraphics[scale=0.45]{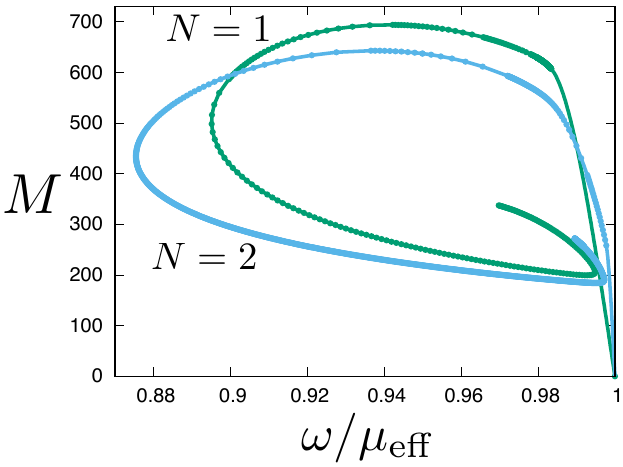}\label{Mj15}
  }
 \caption{The mass of the hairy Kaluza-Klein monopole $M$ as a function of the scalar field frequency $\omega/\mu_\text{eff}$. The numerical data are indicated by dots. The NUT parameter is varied as $N=1/2,1,2$ for panels (a) and (b), and as 
$N=1,2$ for panel (c).
}
 \label{M_vs_omega}
\end{figure}

\begin{figure}[t]
  \centering
\subfigure[$j=k=1/2$]
 {\includegraphics[scale=0.45]{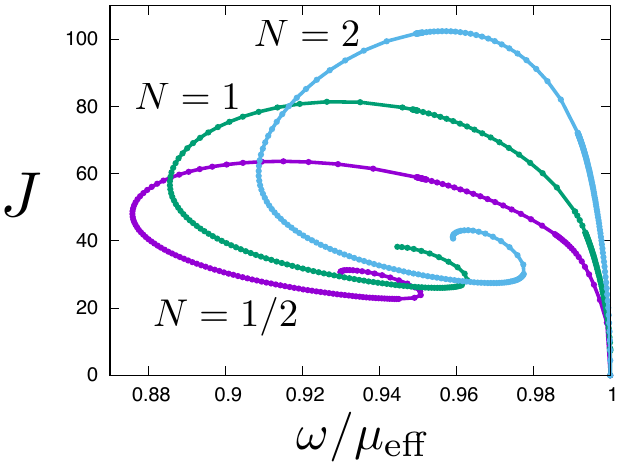}\label{Jj05}
  }
  \subfigure[$j=k=1$]
 {\includegraphics[scale=0.45]{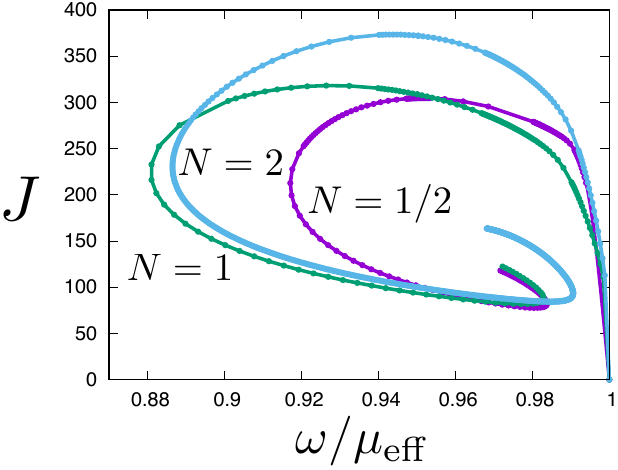}\label{Jj1}
  }
   \subfigure[$j=k=3/2$]
 {\includegraphics[scale=0.45]{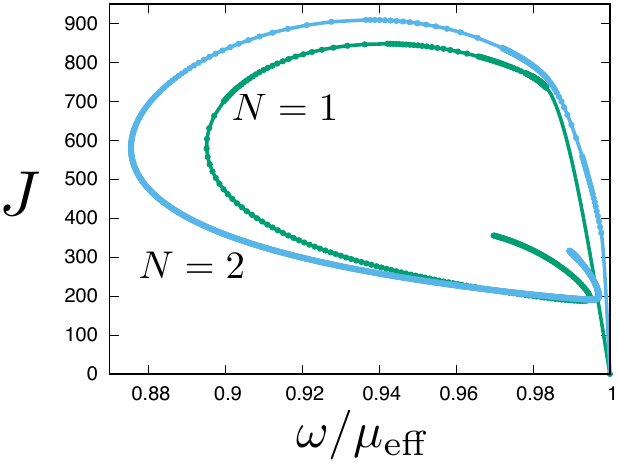}\label{Jj15}
  }
 \caption{The angular momentum of the hairy Kaluza-Klein monopole $J$ as a function of the scalar field frequency $\omega/\mu_\text{eff}$. The numerical data are indicated by dots. The NUT parameter is varied as $N=1/2,1,2$ for panels (a) and (b), and as 
$N=1,2$ for panel (c).
}
 \label{J_vs_omega}
\end{figure}

Fig.~\ref{M_vs_omega} shows the mass $M$ of the hairy Kaluza-Klein monopole as a function of $\omega / \mu_\text{eff}$.\footnote{For $j = k = 3/2$, the numerical construction of the solutions becomes technically difficult for small values of $N$. In particular, for $\omega / \mu_{\text{eff}} = 0.95$, we found solutions only for $N \gtrsim 0.58$. Therefore, we present results only for $N = 1, 2$ in the case of $j = k = 3/2$.}
In the figure, we set $4\pi G_5 = 1$ in Eqs.~\eqref{cc} and \eqref{relativemass}. For $j = k = 1/2$, our results agree with those obtained in~\cite{Brihaye:2023vox}. The results for $j \geq 1$ are new.
For a fixed $N$, the hairy Kaluza-Klein monopole solutions exist only within a finite frequency range $(0 <) \ \omega_{\text{min}} \leq \omega \leq \mu_{\text{eff}}$. The presence of a minimal frequency means that the scalar field is always oscillatory in the hairy Kaluza-Klein monopole spacetime, which is rotating. The mass $M$ and the angular momentum $J$ vanish in the limit $\omega / \mu_{\text{eff}} \to 1$.  The curves exhibit a spiral structure, indicating that, for a given value of $\omega / \mu_\text{eff}$ and $N$, multiple solutions with different $M$ can exist.
Fig.~\ref{J_vs_omega} shows the angular momentum of the hairy Kaluza-Klein monopole as a function of $\omega / \mu_\text{eff}$. The behavior of the angular momentum is similar to that of the mass.

\begin{figure}[t]
  \centering
\subfigure[$N=1/2$]
 {\includegraphics[scale=0.42]{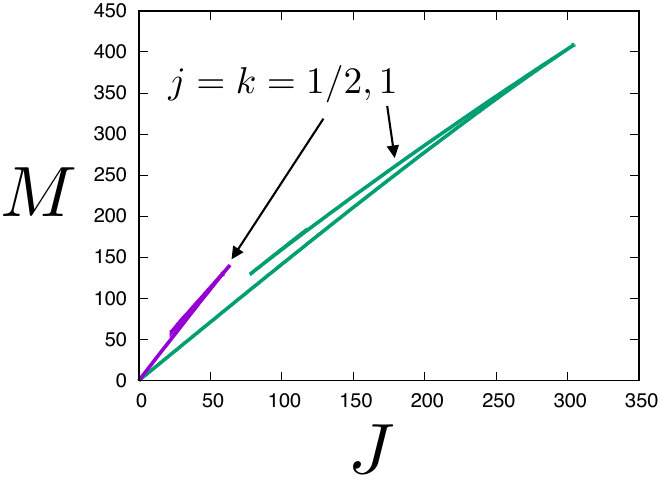}\label{MJ05}
  }
  \subfigure[$N=1$]
 {\includegraphics[scale=0.42]{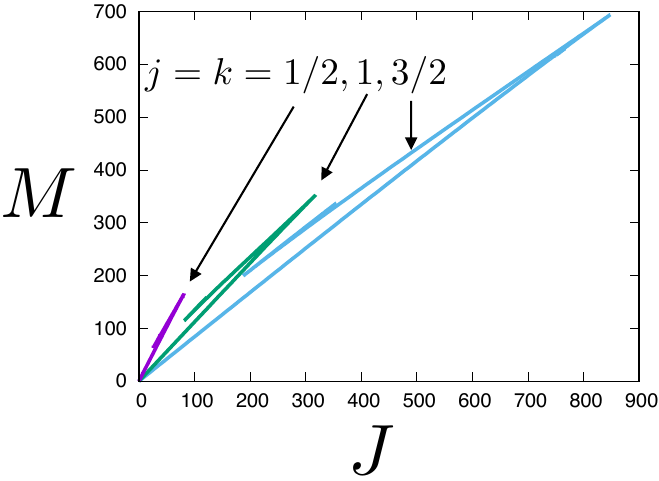}\label{MJ1}
  }
   \subfigure[$N=2$]
 {\includegraphics[scale=0.42]{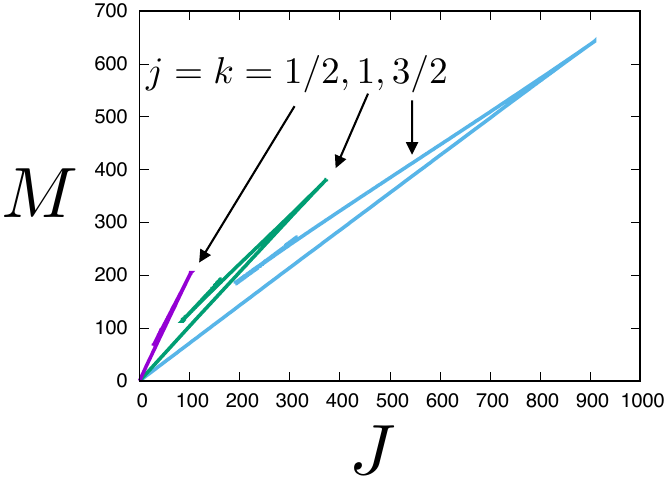}\label{MJ2}
  }
 \caption{The mass of the hairy Kalza-Klein monopole $M$ as the function of its angumar momentum $J$.
 For panels (a), (b) and (c), the NUT parameter is fixed as $N=1/2, 1$ and $2$, respectively.
}
 \label{M_vs_J}
\end{figure}

Fig.~\ref{M_vs_J} shows the mass of the hairy Kaluza-Klein monopoles as a function of the angular momentum $J$. As shown in Figs.~\ref{M_vs_omega} and \ref{J_vs_omega}, $M$ and $J$ exhibit a spiral structure as functions of $\omega$. These functions attain their extrema at same values of $\omega$. Consequently, cusps appear in $M$-$J$ diagrams, as seen in Fig.~\ref{M_vs_J}. 

When $j$ and $N$ are fixed, the mass and angular momentum of the hairy Kaluza-Klein monopole have maximum values. It is observed that these maximum values increase as $j$ increases. This means that by increasing the components of the complex scalar field, one can construct heavier (and more angular momentum) hairy Kaluza-Klein monopole solutions. The ratio of angular momentum and mass $J/M$ is also larger as $j$ increases.

Note that our ansatz is defined with fixed $j$ and a $(2j+1)$-dimensional multiplet.  
However, a solution for a given $j$ remains a valid solution in a theory with more than $2j+1$ multiplets simply by setting the additional components to zero. Hence, solutions with different $j$ can be compared consistently. We can consider an $n$-component scalar multiplet with $n\ge 2j+1$ and impose the $2j+1$ components to be nontrivial but the other $n-2j-1$ components are zero.
For example, the hairy Kaluza-Klein monopole solutions with $j=1/2, 1, 3/2$ can be regarded as solutions in the same theory given by a complex scalar field with four components. In such cases, it is not immediately clear whether some or all solutions are stable. To address this issue, it will be necessary to perform a stability analysis of the solutions.

\section{Conclusion}\label{Conclusion}
We constructed hairy Kaluza-Klein monopoles in five-dimensional Einstein gravity minimally coupled to a massive complex scalar field multiplet. The spacetime has $R_t \times U(2)$ symmetries, and the metric is given in the cohomogeneity-1 form. The key idea for constructing the solution with these symmetries is to introduce synchronized components of the multiplet scalar. In the preceding work \cite{Brihaye:2023vox}, hairy solutions with Kaluza-Klein asymptotics have been considered in the presence of a scalar field doublet. In this paper, we have shown that we can have arbitrary multiplets by expressing the scalar field in terms of the Wigner D-matrices and imposing the ansatz consistent with the symmetry on the scalar multiplet. The solutions were obtained numerically by solving the Einstein-Klein-Gordon equations that were reduced to five coupled simultaneous ordinary differential equations under the ansatz.

In terms of the quantum numbers $j$ and $k$, which are familiar in quantum mechanics where the Wigner D-matrices are utilized and whose range is explicitly given in \eqref{jmk_quantum_numbers}, the preceding work \cite{Brihaye:2023vox} corresponds to the case of $j=k=1/2$ for the doublet scalar, as well as the singlet $j=k=0$. In this paper, we numerically obtained the hairy Kaluza-Klein monopole solutions also for $j=k=1$ and $j=k=3/2$. As shown in Fig.~\ref{M_vs_J}, hairy solutions exist in a larger parameter space of mass $M$ and angular momentum $J$. In particular, we observed the tendency that solutions have a smaller ratio of $M/J$ as the quantum number $j=k$ increases. That is, with a higher scalar multiplet, the hairy Kaluza-Klein monopole rotates more efficiently for a given mass $M$. It will be interesting to examine the bound of $M/J$ in the limit of infinite multiplet $j \to \infty$ and see if there is saturation of a BPS-like condition.

It will be of course a straightforward generalization of this work to also construct the solutions with a black hole horizon for the same $R_t\times U(2)$ symmetric ansatz. Qualitatively, even for higher $j$, black hole solutions will appear in a way similar to the preceding results for the doublet scalar field, whereas explicitly constructing hairy black holes for arbitrary $j$ will significantly expand the parameter space of the existence of the five-dimensional hairy black hole solutions.

\section*{Acknowledgements}
The work of T.\ I.\ was in part supported by Rikkyo SFR. 
The work of K.\ M.\ was supported in part by JSPS KAKENHI Grant No.\ JP21H05186. 

\appendix
\section{Counterterm method for obtaining conserved quantities}\label{Counterterm}

We use the counterterm method to obtain the conserved quantities of spacetime. This is a convenient technique for calculating the quasilocal energy and conserved charges \cite{Brown:1992br}. It has been developed intensively in asymptotically AdS spacetime and is known by the name ``holographic renormalization'' \cite{Balasubramanian:1999re,Kraus:1999di,deHaro:2000vlm,Bianchi:2001kw}, while the method can also be used without the cosmological constant \cite{Kraus:1999di,Mann:2005yr} and in particular in the GPS-monopole spacetime \cite{Mann:2005cx}.

The action including boundary terms is given by
\begin{equation}
    S=\frac{1}{16\pi G_5}\int dx^5 \sqrt{-g}\qty(R-g^{\al\be}\pa_\al\bm{\Pi}^*\cdot\pa_\be\bm{\Pi}-\mu^2\bm{\Pi}^*\cdot\bm{\Pi})+S_B+S_{\text{ct}}\ ,
    \label{counter_S}
\end{equation}
where $S_B$ and $S_{\text{ct}}$ are the Gibbons-Hawking-York boundary term and counterterm, respectively, on the cutoff surface $\partial M$ at large radius which will eventually be sent to asymptotic infinity. The boundary term is given by
\begin{equation}
    S_B=\frac{1}{8\pi G_5}\int_{\pa M}d^4x \sqrt{-\ga}\mathcal{K}\ ,
\end{equation}
where $\ga_{ij}$ and $\mathcal{K}$ denote the induced metric and the extrinsic curvature on $\partial M$.
In this paper, we employ the following counterterm $S_{\text{ct}}$ for the GPS-monopole spacetime as discussed in \cite{Mann:2005cx}:
\begin{equation}
    S_{\text{ct}}=-\frac{1}{8\pi G_5}\int_{\pa M}d^4x \sqrt{-\ga}\sqrt{2\mathcal{R}} \ ,
\end{equation}
where $\mathcal{R}$ is the Ricci scalar on $\partial M$.

From the variation of \eqref{counter_S} with respect to $\gamma_{ij}$, the boundary stress tensor is calculated as
\begin{equation}
\begin{split}
    T_{ij}&=-\frac{2}{\sqrt{-\ga}}\frac{\del S}{\del \ga^{ij}}\\
    &=-\frac{1}{8\pi G_5}\qty{\mathcal{K}_{ij}-\mathcal{K}\ga_{ij}-\Psi\qty(\mathcal{R}_{ij}-\mathcal{R}\ga_{ij})-\gamma_{ij}\gamma^{kl}\Psi_{,kl}+\Psi_{,ij}}\ ,
\end{split}
\end{equation}
where $\Psi=\sqrt{2/\mathcal{R}}$.
The conserved charge is then given by the following integral \cite{Brown:1992br}:
\begin{equation}
Q=\int_\Sigma d^3S \, n^iT_{ij}\xi^j\ ,
\end{equation}
where $n^i$ denotes the unit normal vector of the three-dimensional spacelike hypersurface $\Sigma$ on $\partial M$, and $\xi_i$ a Killing vector.
Using the power series expansion at infinity \eqref{power_series_infinity},\footnote{Note that $g(r) \to 4$ in $r \to \infty$ in our $r$-coordinate as we saw in \eqref{power_series_infinity}, and a bit of thinking is necessary in order to take care of this unfamiliar factor 4.} we obtain the following expression of the mass and angular momentum as conserved charges,\footnote{There is also the tension $T$, constructed from the $\chi\chi$-component of the boundary stress tensor, $T_{\chi\chi}$ \cite{Brihaye:2023vox}. In our notation, it is given by
\begin{equation}
T=-\frac{1}{8NG_5}\qty(\frac{C_\beta}{8N}+16N^2C_f)\ .
\end{equation}
For the GPS monopole, we get $T_\text{GPS}=N/G_5$, and we can check that the Smarr relation is satisfied as $2M_\text{GPS}=LT_\text{GPS}$, where $L=8 \pi N$ is the circumfarence of the extra dimension.}
\begin{align}
    M_{\text{tot}}&=-\frac{\pi}{2G_5}\qty(\frac{C_\beta}{8N}+4NC_f)\ ,\\
    J&=-\frac{8\pi N^3}{G_5}\ .
    \label{cc2}
\end{align}
Here, we use the symbol $M_{\text{tot}}$ for the total mass because it includes the nonzero mass of the GPS monopole. The GPS monopole corresponds to $C_f=0$ and $C_\beta=-64N^3$, and we obtain $M_{\text{GPS}}=4\pi N^2/G_5$. In the main text, when we consider the hairy Kaluza-Klein monopole, we use the mass measured from the GPS monopole, $M=M_\text{tot}-M_\text{GPS}$.

\bibliography{bibl}

\end{document}